\documentclass[11pt]{article}
\usepackage{hyperref}
\pdfoutput=1

\begin{document}

\title{Nanodroplet Impact on Solid Platinum Surface: Spreading and Bouncing}
\author{D.T. Lussier and Y. Ventikos \\
\\\vspace{6pt} Department of Engineering Science, \\ University of Oxford, Oxford, UK}

\maketitle

\noindent The impact of droplets onto solid surfaces is found in a huge variety of natural and technological applications from rain drops splashing on pavement, to material manufacturing by molten droplet deposition.  There is a large range of potential outcomes from any impact event which must be controlled by the appropriate selection of impact conditions if the droplet impact process is to be of use in technological application \cite{Yarin:2006p171}.

Taking inspiration from existing microfluidic technologies (i.e. lab-on-chip), which as shown the appeal of miniaturized fluidic technology,  there is increasing interest in the use of nanodrolets ($D < 100$ nm) for a number of applications such as drug delivery and semiconductor device manufacturing. As the size of the droplet is reduced into the nanoscale, the direct use of previously obtained macroscopic results is not guaranteed. In physical systems of this size, important effects due to the molecular nature of matter, thermal fluctuations and reduced dimensionality can play a critical role in determining system dynamics. 
	
\vspace{0.3cm}

In the linked videos, we have used large-scale, fully atomistic, three-dimensional molecular dynamics (MD) simulation to study the impact of an argon nanodroplet ($D ~ 20$ nm, $54000$ atoms) impact onto a solid platinum surface ($100000$ atoms set in five layers thermostated by Langevin thermostat), using the LAMMPS software package \cite{Plimpton:1995p152}. The argon fluid is modeled using the well known Lennard-Jones (LJ) potential, while the embedded-atom model (EAM) potential is used for the solid platinum.  The argon-platinum cross-interaction is modeled using a Lennard-Jones potential and was tuned to modify the wettability of of the platinum surface from fully wetting to effectively non-wetting.

\vspace{0.3cm}

Impacts at two droplet velocities ($20$ m/s and $100$ m/s) were performed for both wetting and non-wetting surfaces.  A stark difference is seen between the wetting conditions, as at both impact velocities the droplet spread smoothly over the the wetting surface, but bounced back from the non-wetting surface.  It is also worth noting the effect of the increased speed for both wetting conditions.  For impact onto a wetting surface, there is little visible effect of changing impact speed.  At both $20$ and $100$ m/s the droplet is seen to spread out smoothly once brought into contact with the surface at a lateral speed that is comparable to the approach velocity.  In contrast, for the non-wetting impacts, while there is very little droplet deformation for the slower $20$ m/s impact, during the $100$ m/s impact a large amount of droplet deformation is required to create the bouncing outcome.  There is also a clearly visible surface wave resulting in a slight dimpling of top of the droplet that is generated as the droplet mass rushes back inwards after achieving full lateral extent.  This observation highlights the dominance of surface forces over inertia in nanoscale fluidic systems even at moderate droplet velocities.

\vspace{0.3cm}

\noindent The video showing the nanodroplet collisions described can be seen at the following URLs:

	\begin{itemize}
		\item \href{http://ecommons.library.cornell.edu/bitstream/1813/14116/2/Nanodroplet-Pt-Surface-LO.mpg}{Video 1 - Low resolution}
		\item \href{http://ecommons.library.cornell.edu/bitstream/1813/14116/4/Nanodroplet-Pt-Surface-HI.mpg}{Video 2 - High resolution}
	\end{itemize}

\noindent This video has been submitted to the \emph{Gallery of Fluid Motion 2009} which is an annual showcase of fluid dynamics videos.

\vspace{0.3cm}

\noindent The authors would like to thanks and acknowledge the support of the Natural Sciences and Engineering Research Council (NSERC) of Canada for supporting this work.

\bibliographystyle{plain}

\end{document}